\documentclass[aps,prc,preprint,superscriptaddress,showpacs]{revtex4}
\usepackage{graphicx}
\usepackage{bm}
\begin{document}
\title{ Three-$\alpha$-cluster structure of the $0^+$ states in 
$^{12}\mathrm{C}$ and the effective $\alpha - \alpha$ interactions }

\author{S.~I.~Fedotov}
\affiliation{ Bogoliubov Laboratory of Theoretical Physics, Joint
Institute for Nuclear Research, Dubna, 141980, Russia }
\author{O.~I.~Kartavtsev}
\affiliation{ Dzhelepov Laboratory of Nuclear Problems, Joint
Institute for Nuclear Research, 141980, Dubna, Russia }
\author{V.~I.~Kochkin}
\affiliation{Laboratory of Information Technologies, 
Joint Institute for Nuclear Research, Dubna, 141980, Russia}
\author{A.~V.~Malykh}
\affiliation{ Dzhelepov Laboratory of Nuclear Problems, Joint
Institute for Nuclear Research, 141980, Dubna, Russia }
\affiliation{ Physics Department, Novgorod State University, Novgorod
the Great, 173003, Russia }


\begin{abstract}

The $0^{+}$ states of $^{12}\mathrm{C}$ are considered within the framework 
of  the microscopic three-$\alpha$-cluster model. The main attention is paid 
to accurate calculation of the width of the extremely narrow near-threshold 
$0^+_2$ state which plays a key role in stellar nucleosynthesis. 
It is shown that the $0^{+}_2$-state decays by means of the sequential 
mechanism ${^{12}\mathrm{C}} \to \alpha+{^8\mathrm{Be}} \to 3\alpha$. 
Calculations are performed for a number of effective $\alpha - \alpha$ 
potentials which are chosen to reproduce both energy and width of 
$^8\mathrm{Be}$. The parameters of the additional three-body potential 
are chosen to fix both the ground and excited state energies at 
the experimental values. The dependence of the width on the parameters 
of the effective $\alpha - \alpha$ potential is studied in order 
to impose restrictions on the potentials. 

\end{abstract}

\maketitle

\section{Introduction}

The processes with few (three and more) charged particles in the initial 
or final state are complicated phenomena which so far have not been 
completely understood. The main difficulty stems from the necessity 
to describe the continuum wave function of three (or more) charged particles 
(three-body continuum). Reliable description of the continuum three-body 
wave function is of importance for a number of problems in nuclear physics 
and nuclear astrophysics. As the first example one should mention the famous 
nuclear reaction -- formation of the $^{12}\mathrm{C}$ nucleus 
in the triple-$\alpha$ low-energy collisions. This reaction is of key 
importance for stellar nucleosynthesis~\cite{Salpeter52,Hoyle54} 
as a unique possibility for helium burning that allows further synthesis 
of heavier elements. Other interesting examples of the three-body nuclear 
processes are double-proton radioactivity which has been a subject of thorough 
experimental and theoretical investigations during the last years (more 
details can be found in the recent reviews~\cite{Grigorenko01,Grigorenko03}) 
and decay of the long-lived $1^{+}$ state of the $^{12}\mathrm{C}$ 
nucleus~\cite{Fynbo03}. For the problems of this kind, even qualitative 
understanding of the reaction 
mechanism is crucial. In this respect, Coulomb-correlated penetration 
of outgoing particles through a multidimensional potential barrier has been  
considered in Ref.~\cite{Kartavtsev03} thus describing qualitative 
features of multicluster decay of atomic nuclei. 

Of key importance for description of the triple-$\alpha$ reaction are both
the near-threshold three-body resonance ($0_2^+$ state of $^{12}\mathrm{C}$) 
predicted in Ref.~\cite{Hoyle54} as the only explanation for observable 
abundance of elements in the universe and the low-energy $\alpha - \alpha$ 
resonance (the ground state of $^8\mathrm{Be}$). Due to existence 
of these resonances, sufficiently fast helium burning in stars was explained 
by the sequential mechanism $3\alpha \to {^8\mathrm{Be}} + \alpha 
\to {^{12}\mathrm{C}} (0_2^+) \to {^{12}\mathrm{C}} + \gamma$ of the reaction.
Indeed, the predicted $0_2^+$ state of the $^{12}\mathrm{C}$ nucleus was 
observed in the experiments~\cite{Dunbar53,Cooc57} and was studied in 
the later works; in particular, the decay mechanism was a subject of 
investigation in Ref.~\cite{Freer94}. 
The experimental studies must be supplemented by microscopic calculations 
to provide unambiguous determination of the decay mechanism and resonance 
width $\Gamma \sim 8$eV the extremely small on the nuclear scale. 

Besides the resonance triple-$\alpha$ reaction, it is of interest to 
consider in astrophysical applications, as pointed out in 
Ref.~\cite{Cameron59}, 
the non-resonance reaction $3\alpha\to{^{12}\mathrm{C}}$ 
which takes place at low temperatures and high densities. Helium burning 
at such conditions is possible in accretion on white dwarfs and neutron stars. 
The non-resonance reaction was considered in a number of 
papers~\cite{Nomoto85,Langanke86,Fushiki87,Schramm92} based on the model 
assumptions; however, a consistent treatment of the three-body dynamics 
is lacking. In this respect, note that at ultra-low energies any 
approximation can lead to an error of a few orders of magnitude in the 
calculated reaction rate.  

Besides astrophysical applications, studies of the three-$\alpha$  
scattering provides important information about the effective $\alpha - 
\alpha$ interactions which is of interest for the $\alpha$-cluster 
calculations. As the $\alpha$-particle is the most tightly bound nucleus, 
many low-energy nuclear properties can be successfully calculated 
within the framework of the $\alpha$-cluster  
model~\cite{Pichler97,Fedorov96,Filikhin00,Filikhin00a,Descouvemont87,
Kanada-En'yo98}. Generally, the three-body calculations allow one to reduce 
the uncertainty in the two-body potential which can be hardly determined 
only from the two-body data. One of the principal opportunities for 
unambiguous determination of the $\alpha - \alpha$ effective potential 
is to set the calculated width of the $0_2^+$ three-body resonance to its 
experimental value. In addition, one should mention that recently 
the near-threshold $\alpha$-cluster states have attracted 
a special attention in connection with $\alpha$-particle condensation in 
low-density nuclear matter~\cite{Tohsaki01,Yamada04}.

In this paper, properties of the $0^{+}$ states of $^{12}\mathrm{C}$ 
are considered using the $3\alpha$-cluster model with the main emphasis  
on the calculation of the width of the near-threshold $0^+_2$ state. 
Calculations are performed for a number of effective 
$\alpha - \alpha$ potentials which are chosen to reproduce with a good 
accuracy both the energy and the width of the $\alpha - \alpha$ resonance 
(ground state of $^8\mathrm{Be}$). Furthermore, due to the strong 
exponential dependence of the resonance width on the resonance energy, 
calculation of the width makes sense only if the resonance position is fixed. 
For the 3-$\alpha$ resonance under consideration, this requirement 
is satisfied by adjusting the parameters of the additional three-body 
potential which must be introduced to describe the effect of 
$\alpha$-particle distortions near the triple-collision point. 
More precisely, the three-body potential is chosen to fix both the ground 
and excited state energies at the experimental values.  
Following Ref.~\cite{Macek68}, the method of calculation is based 
on the expansion of the total wave function in terms of the eigenfunctions 
on a hypersphere (at a fixed hyper-radius), which allows solving both 
the eigenvalue problem for the ground state and the scattering problem for 
the excited resonance state. 
The eigenfunctions on a hypersphere are calculated by using the variational 
method with a flexible set of trial functions which describe properly 
the three-body wave function both at large and small interparticle distances. 

The present three-body calculation of the near-threshold resonance is 
the first necessary step in the unified treatment of the low-energy 
triple-$\alpha$ reaction. Both the method and the numerical procedure 
can be used 
to calculate the reaction rate at lower energies where the resonance mechanism 
turns to the non-resonance one.

\section{Method}


The present paper is aimed at microscopic description of the low-energy 
scattering of three $\alpha$-particles whose features are to much extent 
determined by the two- and three-body resonances. In this respect, 
the principal problem is reliable calculation of characteristics 
of the extremely narrow near-threshold resonance 
($0_2^+$-state of $^{12}\mathrm{C}$). The $\alpha$-cluster model is used 
that allows for taking account of the most important features of the 
wave function, i.e., 
the three-$\alpha$-cluster and two-cluster $\alpha + ^8\mathrm{Be}$ 
components. All the effects connected with both the internal structure of 
$\alpha$-particles and the identity of nucleons are incorporated in the 
effective 
$\alpha - \alpha$ potential. Besides, the additional three-body potential 
of a simple Gaussian form as in papers~\cite{Fedorov96,Filikhin00} 
is introduced to describe the effects beyond the three-cluster approximation. 
Thus, the model allows description of both the ground and the excited 
$0^+$ states of $^{12}\mathrm{C}$. Considering the challenging problem 
of reliable calculation of the $0_2^+$ resonance width, the effective 
two-body potential should satisfy the restriction that the position 
and width of $^8\mathrm{Be}$ are fixed at the experimental values. 
In a similar way, the three-body potential will be chosen to obtain 
experimental energies for both ground and excited states of $^{12}\mathrm{C}$. 

For the low-energy scattering in question, one should consider only 
the $0^+$ states (the total angular momentum $L = 0$). The units 
$\hbar = m = e = 1$ are used 
throughout the paper unless other is specified. 
The Schr\"odinger equation for three $\alpha$-particles reads
\begin{equation}
\left(-\Delta_{{\mathbf x}_i} - \Delta _{{\mathbf y}_i} +
\sum_{j=1}^3 V(x_j)+V_3(\rho) - E \right)\Psi = 0 
\label{eq}
\end{equation} 
where the scaled Jacobi coordinates are 
${\mathbf x}_i = {\mathbf r}_j-{\mathbf r}_k ,\ 
{\mathbf y}_i = (2{\mathbf r}_i-{{\mathbf r}_j - {\mathbf r}_k})/\sqrt{3}$ and
${\bf r}_i$ is the position vector of the $i$th particle. In the following 
it is convenient to use the hyperspherical coordinates 
$0 \leq \rho < \infty$, 
 $0 \leq \alpha_i$, and $0 \leq \theta_i \leq \pi$ defined as 
\begin{equation}
x_i = \rho \cos\frac{\alpha_i}{2}\ , \quad 
y_i = \rho \sin\frac{\alpha_i}{2}\ , \quad 
\cos\theta_i = \frac{({\mathbf x}_i{\mathbf y}_i)}{x_i y_i} \ .
\label{hcor}
\end{equation}
In the Schr\"odinger equation~(\ref{eq}) the effective two-body potential 
$V(x) $ is a sum of the short-range and Coulomb interactions 
\begin{equation}
V(x) = V_s(x) + \frac{4}{x} 
\label{v2}
\end{equation} 
where the short-range $\alpha - \alpha $ potential 
\begin{eqnarray}
V_s(x) & = & V_r e^{-\mu_{r}^{2} x^2} - V_a e^{-\mu_{a}^{2} x^2}
\label{eff3}
\end{eqnarray} 
is obtained by modification of the Ali-Bodmer potentials~\cite{Ali66}. 
The three-body potential 
\begin{equation}
V_3(\rho) = V_0\,e^{-(\rho/b)^2} 
\label{ef3}
\end{equation}
is chosen as the same function of the hyper-radius $\rho$ as used  
in Refs~\cite{Fedorov96,Filikhin00}. 

\subsection{Eigenfunctions on the hypersphere}

In terms of the hyperspherical variables the Schr\"odinger 
equation~(\ref{eq}) for $L=0$ reads 
\begin{equation} 
\label{eq1}
\left[ -\frac{1}{\rho^5} \frac{\partial}{\partial\rho}
\left( \rho^5 \frac{\partial}{\partial\rho} \right) - 
\frac{4}{\rho^2}\Delta^* +  
\sum_{j=1}^3 V \left( \rho \cos\frac{\alpha_j}{2} \right) + 
V_3\left( \rho \right) - E \right]\Psi = 0
\end{equation}
where 
\begin{equation} 
\label{ham1}
\Delta^*=\frac{1}{\sin^{2}\alpha_i} \left[ 
\frac{\partial}{\partial\alpha_i} 
\left( \sin^{2}\alpha_i \frac{\partial}{\partial\alpha_i} \right)
+\frac{1}{\sin \theta_i} \frac{\partial}{\partial\theta_i}
\left( \sin\theta_i \frac{\partial}{\partial\theta_i} \right)
\right]
\end{equation}
is the grand angular momentum operator up to a constant factor. In order 
to solve both the eigenvalue and scattering problems for Eq.~(\ref{eq1}) 
the total wave function is expanded in a 
series
\begin{equation}
\Psi = \rho^{-5/2}\sum_n f_n(\rho) \Phi_n(\alpha,\theta,\rho) 
\label{coup}
\end{equation} 
on a discrete set of eigenfunctions $\Phi_n$ of the following equation on 
the hypersphere 
\begin{equation}
\label{eqa}
\left[\Delta^* - \frac{\rho^2}{4}\sum_{j=1}^3 
V\left(\rho \cos\frac{\alpha_j}{2}\right) +
\lambda_{n}(\rho)\right]\Phi_n(\alpha, \theta, \rho) = 0 \ ,
\end{equation} 
as proposed in Ref.~\cite{Macek68}.
At each $\rho$ the index $n = 1, 2, 3, \ldots $ enumerates the eigenvalues 
$\lambda_{n}$ in ascending order and the eigenfunctions 
$\Phi_n(\alpha,\theta,\rho)$ are normalized by the conditions 
$\langle\Phi_n|\Phi_m \rangle = \delta_{nm}$ where the notation 
$\langle\cdot|\cdot\rangle$ means the integration over the invariant volume 
on the hypersphere $d\Omega = \sin^2\alpha_i d\alpha_i d\cos\theta_i$.
Due to identity of $\alpha$-particles both the total wave function $\Psi $ and 
the eigenfunctions $\Phi_n(\alpha,\theta,\rho)$ are symmetric under any 
permutation of particles $i, j$, and $k$. 
Given the expansion~(\ref{coup}) of the total wave function, the Schr\"odinger 
equation~(\ref{eq1}) is reduced to the system of hyper-radial equations (HRE)
\begin{eqnarray}
\label{eq3}
&&\left[ \frac{\partial^2}{\partial \rho^2} - \frac{1}{\rho^2}
\left( 4\lambda_n(\rho) + \frac{15}{4} \right) 
+ V_3(\rho) + E\right] f_{n}(\rho)+ \nonumber \\
&&\sum_m \left( Q_{nm}(\rho)\frac{\partial}{\partial\rho} +
\frac{\partial}{\partial\rho}Q_{nm}(\rho) - P_{nm}(\rho) \right) f_m(\rho) = 0 
\end{eqnarray}
where
\begin{eqnarray}
\label{qdef}
Q_{mn}(\rho) & = &\left\langle\Phi_m \biggm| 
\frac{\partial\Phi_n}{\partial\rho}\right\rangle, \\
\label{pdef}
P_{mn}(\rho) & =
&\left\langle\frac{\partial\Phi_m}{\partial\rho} \biggm| 
\frac{\partial\Phi_n}{\partial\rho}\right\rangle. 
\end{eqnarray}


The coefficients $\lambda_n(\rho)$, $Q_{nm}(\rho)$, and $P_{nm}(\rho)$ 
of HRE~(\ref{eq3}) are calculated using the 
variational method for solution of the eigenvalue problem~(\ref{eqa}). The 
variational basis consists of $N$ trial functions $\chi_i$ with the same 
symmetry under permutations of particles as the eigenfunctions 
$\Phi_n(\alpha,\theta,\rho)$. In view of an essentially different structure 
of the eigenfunctions $\Phi_n(\alpha,\theta,\rho)$ at different values 
of $\rho$, it is necessary to use a flexible basis of trial functions 
which allows one to describe the two- and three-cluster structure 
of the wave function in the asymptotic region. 

First of all, the basis contains a set of the symmetric hyperspherical 
harmonics (SHH) $H_{nm}$ which are the eigenfunctions of the operator 
$\Delta^*$, i.~e.,  
\begin{eqnarray}
\left[\Delta^* + K(K+2)\right]H_{nm} =  0 
\label{qpp1}
\end{eqnarray}
where $K = 2n + 3m$, the non-negative numbers $n$ and $m$ enumerate SHH, 
and $2K$ 
is the order of SHH. For explicit construction of SHH it is convenient to use 
another set of the hyperspherical variables $0\leq \xi \leq \displaystyle{
\frac{\pi}{2}}$, 
$ -\pi \leq \varphi_i \leq \pi$~\cite{Smith62,Dragt65} defined by 
\begin{eqnarray}
& \sin\xi  = \sin\alpha_i\sin\theta_i, \nonumber\\
& \cos\xi\cos \varphi_i  =  \cos\alpha_i, \\
& \cos\xi\sin \varphi_i  =  \sin\alpha_i \cos\theta_i \nonumber \ . 
\label{trial2}
\end{eqnarray}
In these variables 
\begin{equation}
H_{nm}(\xi, \varphi)  \sim  \cos^{3 m}\xi P_{n}^{(0,3m)}(\cos2\xi)
T_{3m}(\cos\varphi) \sim
d^{n+\frac{3}{2}m}_{\frac{3}{2}m,\frac{3}{2}m}(2\xi) \cos3m\varphi
\label{trial1}
\end{equation}
where $P_{n}^{(\alpha,\beta)}(x)$ and $T_{n}(x)$ are the Jacobi and Chebyshev 
polynomials and $d_{mk}^{j}(\beta)$ is the Wigner function. The variable $\xi$ 
is invariant under permutations of particles and, therefore, is independent of 
the index $i$ enumerating the Jacobi variables. On the other hand, 
$\varphi_i$ changes to $\varphi_i \pm 2\pi/3$ under the cyclic permutations as 
$\mid\varphi_i-\varphi_j\mid = 2\pi/3$ and $\varphi_i \to - \varphi_i$ under 
the permutation of particles $j$ and $k$. As follows from Eq.~(\ref{trial1}) 
and the above properties of the variables $\xi$ and $\varphi_i$, SHH are 
completely symmetric under any permutation. 

In the numerical calculations, the basis of trial functions contains a set 
of all SHH $\chi_i(\alpha,\theta ) = H_{n_im_i}(\xi, \varphi)$ with those 
indices $n_i$ and $m_i$ for which $K$ does not exceed the maximum value 
$K_{max}$, i.~e., $K_i = 2n_i + 3m_i \leq K_{max}$. One can count that 
the total number of such SHH for which $2n_i + 3m_i\leq K$ equals 
$K(K+6)/12 + 1$ for $K$ being a multiple of 6 and $([K/6] + 1)(K - 3[K/6])$ 
otherwise. Here $[x]$ stands for the entire part of $x$. Usage of SHH 
in the basis of trial functions provides an excellent description of the 
eigenfunctions at small $\rho$, where the kinetic energy term dominates, and 
quite a good description at intermediate $\rho$, where the cluster effects 
still do not dominate. However, the two-cluster component of the wave 
function corresponding to the configuration $\alpha + ^8\mathrm{Be}$ 
can be hardly described by a set of SHH due to rather slow convergence that 
hinders the calculation at sufficiently large $\rho$. 

In order to describe the two-cluster configuration, the basis of trial 
functions should also include the $\rho$-dependent symmetric combinations 
\begin{equation}
\chi_i(\alpha, \theta ) = \sum_{j=1}^{3}\phi_i(\rho\cos\frac{\alpha_j}{2}) 
\label{trial}
\end{equation}
of the two-body functions $\phi_i(x)$ which are chosen to describe 
the wave function of the two-body $\alpha - \alpha$ resonance. More precisely, 
a set of $\phi_i(x)$ includes a few Gaussian functions 
\begin{equation}
\phi_i(x) = \exp{(-\beta_i x^2)} 
\label{tri1}
\end{equation}
which allows the two-body wave function to be described, with properly chosen 
parameters $\beta_i$, within the range of the nuclear potential $V_s(r)$. 
In addition, the function 
\begin{equation}
\phi(x) = x^{1/4}\exp{(-4\sqrt{x}(1 + ax))} 
\label{tri2}
\end{equation} 
is used to describe the two-body wave function in the under-barrier region. 
This latter function is of the asymptotic form of the Coulomb wave function 
which is optionally cut off by the parameter $a$ at large distances. 

Although the eigenvalues $\lambda_n(\rho)$ are directly determined 
in the variational calculation, the coupling terms $Q_{nm}(\rho)$ and 
$P_{nm}(\rho)$ can be hardly determined by means of definitions~(\ref{qdef}) 
and~(\ref{pdef}), which is hindered due to necessity to calculate 
the derivatives $\displaystyle\frac{\partial \Phi_n }{\partial \rho}$. 
For this reason, $Q_{nm}(\rho)$ are calculated by using the exact expression 
\begin{eqnarray}
Q_{mn}(\rho) & = & \frac{3}{4}\left(\lambda_n -\lambda_m
\right)^{-1} \left\langle \Phi_m \left| \frac{q}{\cos
\alpha}+2\rho V_s(\rho \cos\frac{\alpha}{2})+ \rho^2
\frac{\partial V_s(\rho \cos\frac{\alpha}{2})}{\partial \rho}
\right|\Phi_n \right\rangle 
\label{qp}
\end{eqnarray}
which is derived by differentiating the eigenvalue equation~(\ref{ham1}) 
with respect to $\rho$ and projecting the result on the function $\Phi_m$. 
Furthermore, 
$P_{nm}(\rho)$ are calculated by using the exact sum rule 
$\mathrm{P} = -\mathrm{Q}^2$ for the matrices $\mathrm{P}$ and $\mathrm{Q}$,  
which leads to the approximation 
\begin{eqnarray}
P_{mn}(\rho) & = & \sum_{l=1}^{N}Q_{ml}(\rho)Q_{nl}(\rho) 
\label{qp1}
\end{eqnarray}
on the limited basis of $N$ trial functions. 

\subsection{Boundary conditions and characteristics of $^{12}\mathrm{C}$ 
states}

Properties of the ground $0_1^+$ state and the excited $0_2^+$ resonance  
are determined by solving the eigenvalue problem (at $E<0$) and 
scattering problem (at $E>0$) for HRE~(\ref{eq3}), respectively. 
Denote the hyper-radial 
functions as $f_n^{(1)}(\rho)$ for the ground state and $f_n^{(E)}(\rho)$  
for the scattering problem at energy $E$. According to~(\ref{coup}), all 
these functions satisfy the zero boundary conditions at $\rho=0$. 
The square integrable solution of HRE~(\ref{eq3}) satisfying the condition 
\begin{equation}
\sum_{n}\int\limits_0^{\infty}\left|f^{(1)}_n(\rho)\right|^2 d\rho = 1
\label{nor1}
\end{equation}
unambiguously determines the energy $E_{gs}$ and the wave function of the 
ground $0_1^+$ state. 

The position $E_r$ and width $\Gamma$ of the near-threshold $0_2^+$ resonance 
are calculated by solving HRE~(\ref{eq3}) with the asymptotic boundary 
conditions corresponding to the ingoing wave in the first channel, 
i.~e., $f_1^{E}(\rho)$ is a sum of the ingoing and outgoing waves in 
the effective potential $U_1(\rho) = 
\displaystyle\frac{1}{\rho^2}\left[4\lambda_1(\rho) + 15/4\right]$. More 
precisely, the asymptotic boundary conditions are imposed near the turning 
point $\rho_t$ of the first-channel effective potential defined 
by the condition $U_1(\rho_t) = E$. As expected, calculations reveal (see also 
Fig.~\ref{fig1}) that at $\rho \sim \rho_t$ the effective potential 
$U_1(\rho)$ is to a good approximation expressed as
\begin{equation}
\label{ef1}
U_{1}(\rho) \approx E_{2\alpha}+\frac{\tilde q}{\rho} 
\end{equation}
where $E_{2\alpha}$ is the energy of the two-body resonance (the ground 
state of ${^{8}\mathrm{Be}}$) 
and the Coulomb parameter $\tilde q=16/\sqrt{3}$. In fact, r.h.s. of 
Eq.~(\ref{ef1}) is the energy of the two-cluster system 
$\alpha+{^{8}\mathrm{Be}}$ at a large fixed hyper-radius $\rho$. 
For the scattering at energy above the two-body resonance ($E > E_{2\alpha}$), 
in view of expression~(\ref{ef1}), the first channel hyper-radial function can 
be written as 
\begin{equation}
f_1^{(E)}(\rho) \sim F_0(\eta, k\rho) + \tan\delta_{E} G_0(\eta, k\rho)
\label{limeq2}
\end{equation}
in the range of hyper-radius values $\rho \sim \rho_t$. 
Here the wave number in the first channel $k = \sqrt{E - E_{2\alpha}}$, 
$F_0(\eta,k\rho)$ and $G_0(\eta,k\rho)$  are the Coulomb functions with 
the parameter $\eta = 8/(\sqrt{3}k)$, and $\delta_E$ is the scattering phase 
shift. Due to strong repulsive potentials $U_n(\rho)$ for $n \ge 2$ the 
outgoing waves in the upper channels are negligible at small energies 
$E \le 1$MeV. This allows the zero boundary conditions 
\begin{equation}
f_n^{(E)}(\rho) = 0 
\label{limeq1}
\end{equation}
to be imposed at some value of the hyper-radius $\rho > \rho_t$ for all 
$n \ge 2$. The resonance position $E_r$ and width $\Gamma$ as well as the 
non-resonant phase shift $\delta_{bg}$ are defined by fitting the calculated 
near-resonance phase shift $\delta_E$ to the Wigner dependence on energy
\begin{equation} 
\label{res1}
\cot(\delta_E - \delta_{bg}) = \frac{2}{\Gamma}(E-E_r)\ .
\end{equation}

In the following description of the $0_2^+$ state it is suitable to treat 
the ultra-narrow resonance as a true bound state with the wave function 
$f^{(2)}_n(\rho) \sim f^{(E_r)}_n(\rho)$ corresponding to the scattering 
solution at the resonance energy $E_r$ and normalized on the finite interval 
$0 \le \rho \le \rho_t $ by the condition 
\begin{equation}
\sum_{n}\int\limits_0^{\rho_t}\left|f^{(2)}_n(\rho)\right|^2 d\rho = 1 \ . 
\label{nor2}
\end{equation} 
It is of interest to determine also the root-mean-square (RMS) radii 
\begin{equation}
R^{(i)} = \frac{1}{N_t}\sum_{k}^{N_t}\langle\Psi^{(i)}\left| ({\mathbf r}_k - 
{\mathbf R_{cm}})^2 \right|\Psi^{(i)}\rangle 
\label{rms1}
\end{equation}
of the ground ($i = 1$) and excited ($i = 2$) states and the monopole 
transition matrix element 
\begin{equation}
M_{12} = \sum_{k}^{N_p}\langle\Psi^{(1)}\left| ({\mathbf r}_k - 
{\mathbf R_{cm}})^2 \right|\Psi^{(2)} \rangle \ . 
\label{tranme}
\end{equation} 
A sum is taken over $N_t$ nucleons in~(\ref{rms1}) and over $N_p$ protons 
in~(\ref{tranme}) and ${\mathbf R_{cm}}$ is the center-of-mass position 
vector. Following the definitions~(\ref{rms1}) and~(\ref{tranme}), 
in the three-$\alpha$-particle model one obtains the expressions 
\begin{equation}
\label{rms2}
R^{(i)} = \sqrt{R^2_{\alpha}+\frac{1}{6} \bar\rho^2_i } \ , 
\end{equation}
\begin{equation}
\label{tranme2}
M_{12} = \sum_{n}\int\limits_0^{\rho_t}f^{(2)}_n(\rho)f^{(1)}_n(\rho)\rho^2 
d\rho 
\end{equation}
where $R_{\alpha} = 1.47$fm is the RMS radius of the $\alpha$-particle and 
the RMS value of the hyper-radius $\bar\rho^2_i$ for the $i$th state 
in the three-body model is defined by
\begin{equation}
\bar\rho^2_i = \sum_{n}\int\limits_0^{\infty}\left|
f^{(i)}_n(\rho)\right|^2\rho^2 d\rho \ . 
\label{rmsa}
\end{equation}

\section{Numerical results}

Calculations have been performed with four $\alpha - \alpha$ potentials  
which are obtained by modification of potentials (a) and (d) from 
Ref.~\cite{Ali66}. The parameters of the potentials have been chosen 
to reproduce the experimental value of the $\alpha - \alpha$ resonance 
($^8\mathrm{Be}$) energy $E_{2\alpha} = 91.89$keV whereas resonance widths 
have been allowed to vary within the experimental uncertainty $\pm 1.7$eV. 
The parameters of the potentials and the corresponding $\alpha - \alpha$ 
resonance widths $\gamma$ are presented in Table~\ref{tab1}. Only the 
strengths $V_r$ and $V_a$ of the repulsive and attractive parts are varied 
for potentials 1 -- 3, while the parameters $\mu_r$ and $\mu_a$ 
are modified for ``harder'' potential 4. 
\begin{table}[htb]
\caption{Parameters of the short-range $\alpha - \alpha$ potential 
$V_s$~(\protect\ref{eff3}) and the corresponding widths $\gamma$ of the 
$\alpha - \alpha$ resonance (the ground state of ${^{8}\mathrm{Be}}$). }
\label{tab1}
\begin{tabular}{cccccc}
\multicolumn{1}{c}{  }&\multicolumn{1}{c}{$V_r
$(MeV)}& \multicolumn{1}{c}{$\mu_r$(fm$^{-1}$)}&
\multicolumn{1}{c}{$V_a$(MeV)}& \multicolumn{1}{c}{$\mu_a$
(fm$^{-1}$)}&
\multicolumn{1}{c}{$\gamma $(eV)}\\
\hline
 $1$     & 82.563   & 1/1.53    & 26.1  & 1/2.85 &  6.80 \\
 $2$    & 279.206  & 1/1.53    & 40    & 1/2.85 &  8.53 \\
 $3$   & 20.012   & 1/1.53    & 16.5  & 1/2.85 &  5.11 \\
 $4$   & 197.680  &  0.7      & 80    & 0.475  & 5.10  \\
\end{tabular}
\end{table} 
This choice of the potentials makes it possible to study the dependence of 
the three-body characteristics on the shape of the two-body potential. 

The energies of the ground and excited states $E_{gs}$ and $E_r$, width 
of the excited state $\Gamma$, RMS radii $R^{(i)}$, and monopole 
transition matrix element $M_{12}$ for the three-$\alpha$ system 
are calculated, as discussed in the preceding section, by numerical solution 
of HRE~(\ref{eq3}) with boundary conditions (\ref{limeq2}) 
and (\ref{limeq1}). For the eigenvalue problem, numerical integration 
in the interval $0 < \rho \leq 25$fm provides the relative accuracy 
not worse than $10^{-6}$ for the calculated binding energy. For the scattering 
problem, numerical integration is carried out in the interval 
$0 \leq \rho \leq \rho_{max}$ so that $\rho_{max}\sim 50$fm is chosen 
beyond the turning point $\rho_t \approx 45$fm for the first-channel effective 
potential $U_1(\rho)$. Choice of the parameters of the numerical integration 
does not affect the final results, i.~e., the accuracy of the calculated 
$E_{gs}$, $E_r$, $\Gamma$, $R^{(i)}$, $M_{12}$ depends merely on the accuracy 
of the numerical calculation for $\lambda_n(\rho)$, $P_{mn}(\rho)$, and 
$Q_{mn}(\rho)$. In every calculation, i.~e., for each two-body potential 
and each number of HRE $N_1$, the parameters $V_0$ and $b$ of the three-body 
potential $V_3(\rho)$ are chosen to fix the calculated $E_r$ and $E_{gs}$ 
at the experimental values $E_r = 0.3795$MeV and $E_{gs} = -7.2747$MeV. 

\begin{figure}[htb]
\caption{Three lowest effective potentials $U_n(\rho)$. Shown are also 
the asymptotic two-cluster dependence 
$U_{1}(\rho) \approx E_{2\alpha} + \displaystyle{\frac{\tilde q}{\rho}}$ 
(thin dashed line) 
and the energy of the $0_2^{+}$ state $E = 0.3795$MeV (thin horizontal line).}
\includegraphics[width=0.6\textwidth]{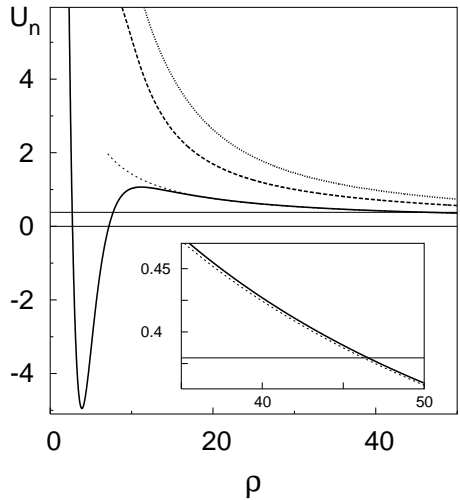}\\
\label{fig1}
\end{figure}

For the calculation of the HRE coefficients $\lambda_n(\rho)$, $P_{mn}(\rho)$, 
and $Q_{mn}(\rho)$ by the variational method, two different sets of trial 
functions were used in the interval $\rho \leq 20$fm and in the 
asymptotic region $\rho > 20$fm. 
At $\rho \leq 20$fm the basis of trial functions contains 147 SHH, 
which corresponds to  $K_{max} = 39$. Recall that convergence 
with a number of SHH becomes very slow with increasing hyper-radius. 
For this reason, at $\rho>20$fm the basis of trial functions contains 
$108$ SHH ($K_{max} = 33$) and four trial functions which 
describe the cluster configuration  $\alpha + {^8\mathrm{Be}}$, namely, 
three functions of the form~(\ref{tri1}) and one function of the 
form~(\ref{tri2}). The $\rho$-independent parameters $\beta_i$ 
in functions~(\ref{tri1}) were determined by minimizing the
first eigenvalue $\lambda_1(\rho)$. The calculated $\lambda_n(\rho)$, 
$P_{mn}(\rho)$, and $Q_{mn}(\rho)$ practically do not depend on the basis 
of trial functions near $\rho = 20$fm, which allows matching the results
of calculation with two different basises.  The three lowest effective 
potentials $U_n(\rho) 
= \displaystyle\frac{1}{\rho^2}[4\lambda_n(\rho) + 15/4] + P_{nn}(\rho)$
($n=1-3$) for two-body potential 3 are depicted for illustration 
in Fig.~\ref{fig1}. 

\begin{figure}[thb]
\caption{The first eigenfunction $\Phi_1(\xi, \varphi, \rho)$ at different 
$\rho$. }
\includegraphics[width=0.45\textwidth]{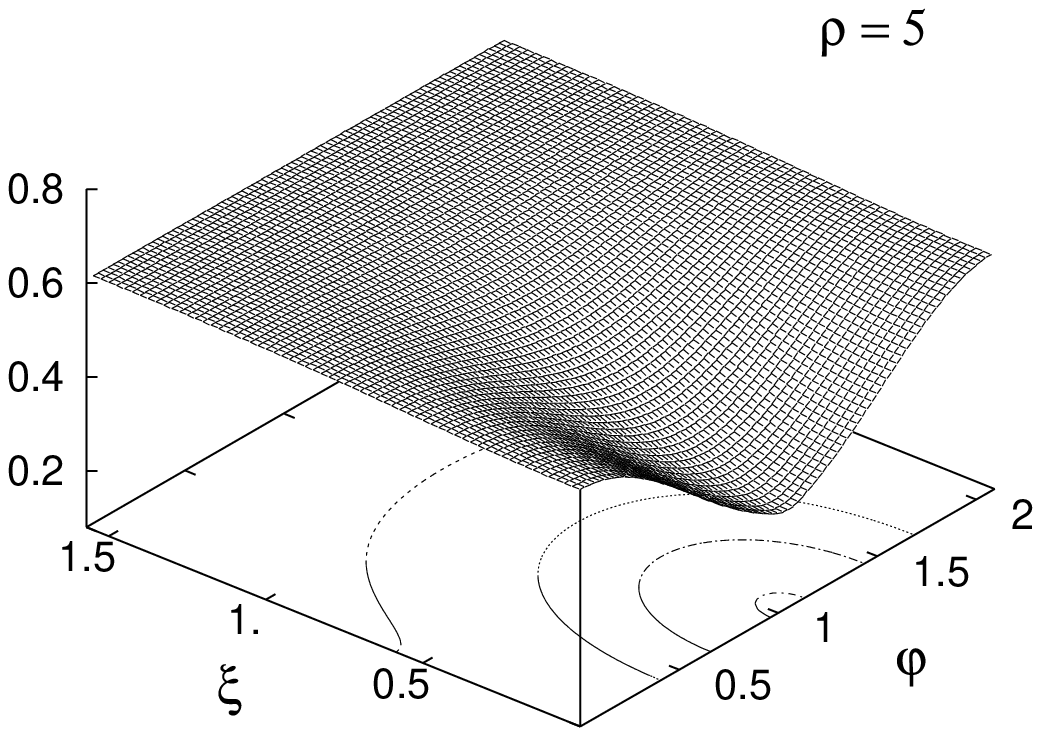}
\includegraphics[width=0.45\textwidth]{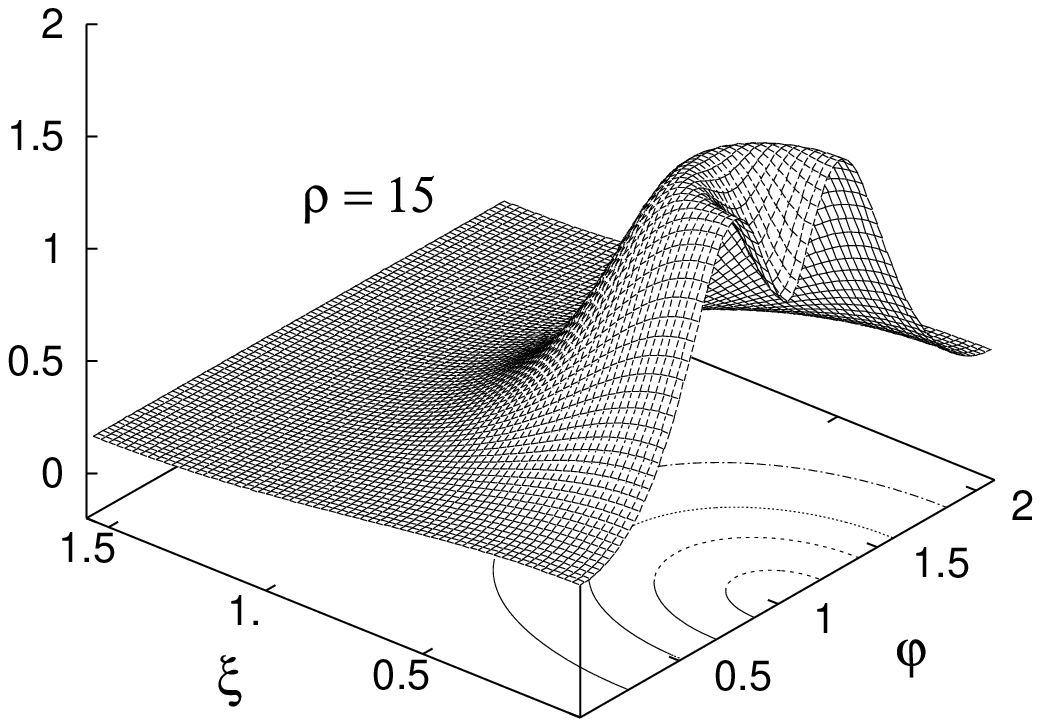}
\includegraphics[width=0.45\textwidth]{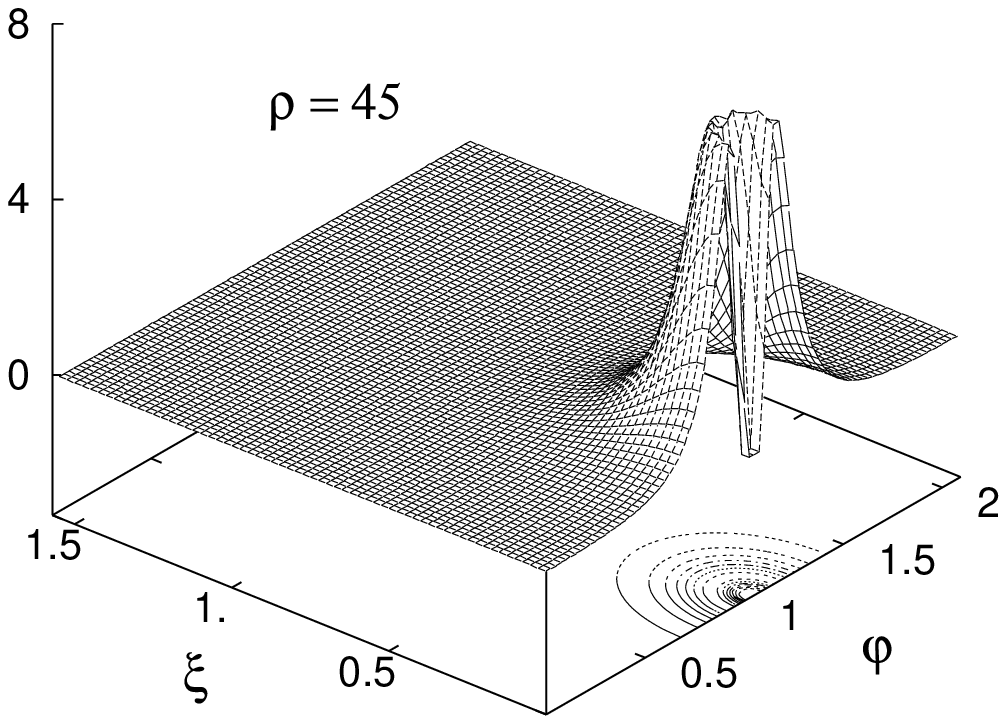}\\
\label{fig2}
\end{figure}

The detailed information on the three-body system can be obtained by 
considering the eigenfunctions on a hypersphere $\Phi_n(\xi, \varphi, \rho)$. 
Due to symmetry, $\Phi_n(\xi, \varphi, \rho)$ are periodic functions 
of the variable $\varphi$ with the period $2\pi/3$. As the main contribution 
to the total wave function comes from the first term ($n = 1$) of 
expansion~(\ref{coup}), the first eigenfunction $\Phi_1(\xi, \varphi, \rho)$ 
practically determines the structure of the system. 
For illustration, three-dimensional plots of the first eigenfunction
$\Phi_1(\xi, \varphi, \rho)$ (for two-body potential 3) are shown in
Fig.~\ref{fig2} at three values of the hyper-radius $\rho=5$fm, $\rho=15$fm, 
and $\rho=45$fm that correspond, as seen in Fig.~\ref{fig1}, to the minimum, 
the maximum, and the turning point of $U_1(\rho)$ at $E=0.3795$MeV. 
For all values of the hyper-radius, $\Phi_1(\xi, \varphi, \rho)$ is small at  
the zero distance between a pair of $\alpha$-particles, i.~e., at the point 
$\xi = 0$, $\varphi=\pi/2$ at the hypersphere, because of the strong 
repulsive term in the short-range potential~$V_s$. 
At sufficiently large $\rho$ (near the turning point), 
$\Phi_1(\xi, \varphi, \rho)$ exhibits a specific structure concentrated 
in the region where the two-body potential is attractive, i.~e., around the 
point $\xi = 0$, $\varphi=\pi/2$. This structure 
corresponds to the two-cluster configuration $\alpha + {^8\mathrm{Be}}$. 
For smaller values of $\rho$ (as seen in Fig.~\ref{fig2} 
at $\rho=15$fm) the two-cluster structure widens; besides, 
visible values of $\Phi_1(\xi, \varphi, \rho)$ appear both at the point 
$\xi = \pi/2$, which corresponds to the configuration of the equilateral 
triangle, and near the  point $\xi = 0$, $\varphi = 0$ (or $\xi = 0$, 
$\varphi = 2\pi/3$), which corresponds to the linear configuration. 
Next, at small $\rho$ near the minimum of $U_1(\rho)$, 
the most important is the triangle configuration with a noticeable weight 
of the linear configuration and without any trace of the two-cluster 
structure. 

The accuracy of the calculation is estimated by observing the convergence with 
increasing number of SHH. As a result, the accuracy of the most significant 
effective potential $U_1(\rho)$ turns out to be not worse than $1$eV in 
the entire interval $0 < \rho \leq 20$fm and is much better for smaller 
$\rho$. In a similar way, 
the relative accuracy of $P_{mn}(\rho)$ and $Q_{mn}(\rho)$ is better than 
$10^{-5}$ in the same interval $0 < \rho \leq 20$fm. As the hyper-radius 
increases beyond $\rho = 20$fm, the accuracy of the calculation decreases due 
to a more complicated structure of the eigenfunctions 
$\Phi_1(\xi, \varphi, \rho)$. Nevertheless, as shown in Fig.~\ref{fig1}, 
the effective potential $U_1(\rho)$ is in good agreement with the asymptotic 
dependence~(\ref{ef1}) in the interval $40\leq \rho \leq 60$, thus pointing 
to the sufficiently high accuracy. In particular, for all the potentials used, 
a fit of $U_1(\rho)$ to Eq.~(\ref{ef1}) gives the values of the Coulomb 
parameter $\tilde q$ which differ from the expected value $\tilde q = 13.3$MeV 
by less than $0.14$MeV$\cdot$fm. The fitted values of $E_{2\alpha}$ differ 
from the experimental energy of ${^8\mathrm{Be}}$ by less than $0.004$MeV 
for potentials 1 -- 3 and by about $0.009$MeV for potential 4. 

Note that representation~(\ref{ef1}) of the effective potential $U_1(\rho)$ in 
terms of the energy of the two-cluster system $\alpha+{^{8}\mathrm{Be}}$
confirms the sequential mechanism of $0_2^{+}$ state decay with formation of 
$\alpha + {^8\mathrm{Be}}$ at the first step. Also, this conclusion 
follows from the genuine two-cluster form of the first-channel eigenfunction
$\Phi_1(\xi,\phi,\rho)$ at $\rho \approx \rho_t$. As shown in Fig.~\ref{fig2} 
at $\rho = 45$fm, $\Phi_1(\xi,\phi,\rho)$ practically coincides with 
the symmetric combination of the ${^{8}\mathrm{Be}}$ wave functions. 
Thus the total width is determined by the two-cluster decay width. 

The calculated $\Gamma$, $R^{(i)}$ ($i=1,2$), $M_{12}$, 
and the parameters of the three-body potential $V_0$ and $b$ are presented 
in Table~\ref{tab2} for four two-body potentials and different number of 
HRE $N_1$.
\begin{table}[htb]
\caption{Characteristics of the $0^+$ states of three $\alpha$-particles 
for four $\alpha$-$\alpha$ potentials calculated with $ N_1 $ HRE. 
The width of the excited state $\Gamma$, the RMS radii $R^{(1)}$ and 
$R^{(2)}$, the monopole transition matrix element $M_{12}$, and 
the parameters $V_0$ and $b$ of the three-body potential are given. 
The results of the calculations~\protect\cite{Fedorov96, Filikhin00} 
and the experimental 
values are given in the last rows. 
\label{tab2}} 
\begin{tabular}{cccccccc}
 & $ N_1 $ & $ \Gamma $(eV) & $R^{(1)}$(fm) & $ R^{(2)} $(fm) & 
$ M_{12}$ (fm$^2$) & $ V_0 $(MeV) & $ b $(fm) \\
\hline
     & 1 & 25 & 2.56 & 4.1 & 9.52 & -21.998 & 4.8993 \\
 $1$ & 2 & 20 & 2.55 & 4.1 & 9.08 & -23.993 & 4.6608 \\
     & 3 & 19 & 2.55 & 4.0 & 9.03 & -24.048 & 4.6530 \\
     & 1 & 37 & 2.82 & 4.4 & 10.3 & -29.291 & 5.1213 \\
 $2$ & 2 & 27 & 2.77 & 4.2 & 9.00 & -39.124 & 4.5223 \\
     & 3 & 26 & 2.77 & 4.2 & 8.85 & -39.947 & 4.4858 \\
     & 1 & 13 & 2.25 & 3.6 & 8.23 & -15.812 & 4.4506 \\
 $3$ & 2 & 11 & 2.24 & 3.6 & 8.15 & -16.059 & 4.3964 \\
     & 3 & 10 & 2.24 & 3.5 & 8.14 & -16.066 & 4.3942 \\
     & 1 & 21 & 2.51 & 4.1 & 8.74 & -14.548 & 5.7248 \\
 $4$ & 2 & 16 & 2.49 & 4.0 & 8.36 & -15.129 & 5.4595 \\
     & 3 & 15 & 2.49 & 4.0 & 8.21 & -15.285 & 5.4118 \\
 ~\cite{Fedorov96} &  & 20 & 2.36 &  & 6.54 & -96.8 & 3.9$/\sqrt{2}$ \\
 ~\cite{Filikhin00} &  & 1300 & 2.47 &  & 8.36 & -23.32 
& 3.795$\cdot\sqrt{2}$\\
 $Exp.$ & & 8.5$\pm 1.0$& 2.47 &  & 5.7 &  &  
\end{tabular}
\end{table}
These results are compared with the experimental values and 
calculations~\cite{Fedorov96, Filikhin00} (the factor $\sqrt{2}$ in the last 
column appears due to different definitions of $\rho$ in these papers). 
As shown in Table~\ref{tab2}, the convergence in a number of HRE is 
sufficiently fast and solution of three HRE ($N_1=3$) allows the resonance 
width to be determined with the accuracy about 1$eV$. The final accuracy 
of $ \Gamma $ depends mainly on the accuracy of the variational 
calculation and can be estimated as a few $eV$. 
The values of the parameter $b$ presented in Table~\ref{tab2} are quite 
reasonable since they are in agreement with the value 
$\rho=2\sqrt{2}R_{\alpha}\approx 4.16$fm, which corresponds to triple 
collision of three hard spheres with radii $R_{\alpha}$. 

\section{Discussion and conclusion}

The main result of the present calculation is accurate determination of an 
extremely narrow width $\Gamma$ of the three-body resonance (the excited 
$0^{+}_2$ state of ${^{12}\mathrm{C}}$). As shown in Tables~\ref{tab1} 
and~\ref{tab2}, the width $\Gamma$ of the three-body resonance 
depends on the underlying two-body $\alpha-\alpha$ potential and 
essentially increases with increasing width $\gamma$ of the two-body
$\alpha-\alpha$ resonance. In fact, variation of $\gamma$ within the limits 
of the experimental uncertainty gives rise to a change of $\Gamma$ by a factor 
2.6; therefore, the precise value of $\gamma$ is necessary for determination 
of $\Gamma$. Note that for all the potentials considered the calculated  
$\Gamma$ overestimate the experimental value $\Gamma_{exp} = 8.5\pm 1$eV 
by a factor $1.2-3.1$. However, the difference of the calculated width 
for potential 3 and the experimental value is of order of the theoretical and 
experimental uncertainties. In addition, comparing the results for potentials 
3 and 4, one may conclude that width $\Gamma$ of the three-body resonance 
depends on the potential shape (on the parameters $\mu_r$ and $\mu_a$). 
It should be emphasized that dependence of $\Gamma$ on the parameters 
of the $\alpha-\alpha$ interaction is rather complicated due to addition 
of the three-body potential $V_3(\rho)$ which is chosen to fix $E_{gs}$ 
and $E_r$ at the experimental values. Nevertheless, the effect of $V_3(\rho)$ 
is not overwhelming since $\Gamma$ is mainly determined by penetrability of 
the potential barrier and all the three-body potentials used in the present 
calculation are rapidly decreasing with increasing $\rho$ in the barrier 
region $10$fm$< \rho <45$fm. 

Similar dependence of $\gamma$ is revealed for the ground-state RMS radius 
$R^{(1)}$. It is interesting that the calculated $R^{(1)} = 2.55$fm is close 
to the experimental value $R_{exp}^{(1)} = 2.47$fm for potential 1, 
for which the two-body width $\gamma$ coincides with the most probable 
experimental value $\gamma_{exp} = 6.8$eV. Similar to $\Gamma$, the calculated 
$R^{(1)}$ essentially depends on the parameters $\mu_r$ and $\mu_a$. Both the  
RMS radius $R^{(2)}$ of the exited states and the monopole transition matrix 
element $M_{12}$ are weakly dependent on the parameters of the two-body 
potential. For all the potentials, the calculated values of $M_{12}$ 
significantly overestimate the experimental value $5.7$fm$^{2}$ that clearly 
deserves further investigation.
 
Comparison with the previous microscopic calculation~\cite{Fedorov96} shows 
that the accuracy of the present calculation is much better than in 
Ref.~\cite{Fedorov96} whereas the methods of calculation are similar to each 
other. Note that the  $\alpha-\alpha$ potential in Ref.~\cite{Filikhin00}  
was chosen to fix the RMS radius of the ground state at the 
experimental value $R_{exp}^{(1)}=2.47$fm. With this potential, the energy of 
the three-body resonance is misplaced by $0.47$MeV. This is essentially 
above the experimental values and leads to an unreliably large resonance 
width. 

The above discussion allows the conclusion that calculation of three-body 
observables can be used to impose restriction on the effective two-body 
$\alpha-\alpha$ potential. In the future one should look for a possibility 
of reproducing the experimental values by a refined choice of potentials, 
in particular, by using more complicated three-body potentials. 

In conclusion, the three-$\alpha$-cluster model is used to calculate the 
characteristics of the  $0^{+}$ states in the ${^{12}\mathrm{C}}$ nucleus. 
In particular, the width of the extremely narrow threshold  $0^{+}_2$ state 
is calculated with a good accuracy. The dependence of the width on the 
parameters of the $\alpha - \alpha$ potential is studied. It is proposed 
to use the calculation of the width for selection of $\alpha-\alpha$ 
potentials. It is directly shown in the three-body calculation that the 
$0^{+}_2$ state decays by means of the sequential mechanism 
(${^{12}\mathrm{C}} \to \alpha+{^8\mathrm{Be}} \to 3\alpha$). 
This conclusion is in agreement with the experiment~\cite{Freer94} in which
the rate of the direct decay ${^{12}\mathrm{C}} \to 3 \alpha$ is estimated 
to be less than 1\% of the total rate. 
The calculation of the near-threshold $3 \alpha$-resonance can be considered 
as a part of the general study of both resonant and non-resonant reaction 
$3\alpha\to{^{12}\mathrm{C}}$ at low energy. 
The present approach is promising for calculation of the triple-$\alpha$ 
reaction rate at low energy below the three-body resonance. This provides 
an opportunity for unified treatment of the crossover from the resonant to 
the non-resonant mechanism of the reaction. 

\bibliography{3alpha}

\end{document}